\documentclass{article}

\usepackage{arxiv}
\usepackage{times}
\usepackage{epsfig}
\usepackage{graphicx}
\usepackage{xspace,flushend}
\usepackage{amsmath}
\usepackage{amssymb}

\DeclareRobustCommand\onedot{\futurelet\@let@token\@onedot}
\def\@onedot{\ifx\@let@token.\else.\null\fi\xspace}

\def\etal{\emph{et al}\onedot}

\title{Machine Learning Method for Light Field Refocusing}

\author{Eisa Hedayati\\
        New York University Grossman School of Medicine\\
        {\tt\small eisa.hedayati@nyu.edu}
        \And
        Timothy C. Havens\\
        Michigan Technological University\\
        {\tt\small thavens@mtu.edu}
        \And
        Jeremy P. Bos\\
        Michigan Technological University\\
        {\tt\small jpbos@mtu.edu}
}

\begin{document}


\maketitle

\begin{abstract}
    Light field imaging introduced the capability to refocus an image after capturing. Currently there are two popular methods for refocusing, shift-and-sum and Fourier slice methods. Neither of these two methods can refocus the light field in real-time without any pre-processing. In this paper we introduce a machine learning based refocusing technique that is capable of extracting 16 refocused images with refocusing parameters of $\alpha=0.125,0.250,0.375,...,2.0$ in real-time. We have trained our network, which is called RefNet, in two experiments. Once using the Fourier slice method as the training---i.e., ``ground truth''---data and another using the shift-and-sum method as the training data. We showed that in both cases, not only is the RefNet method at least $134\times$ faster than previous approaches, but also the color prediction of RefNet is superior to both Fourier slice and shift-and-sum methods while having similar depth of field and focus distance performance.
\end{abstract}

\section{Introduction}
In order to have a sharp image with the desired information about the capture scene, the choice of aperture size and focus point are very important. But if one can obtain a complete \emph{light field} (LF) from the scene, one can synthesize the aperture of choice at any desired focus point by post-processing. By summing the views inside the aperture of choice, one can quickly obtain a synthetic-aperture image from the LF. In the case of a plenoptic camera, one captures two extra dimensions in addition to the regular two spatial dimensions. This four dimensional LF enables limited refocusing and synthetic-aperture \emph{depth of field} (DoF) images by post-processing, which are not available in conventional cameras \cite{adelson1991plenoptic,Levoy1996,ng2005}. The shift-and-sum method \cite{Levoy_SS} and Fourier refocusing \cite{Ng2005Fourier} are two widely used methods for LF refocusing.

The shift-and-sum refocusing method requires 4-dimensional integration for each new refocused image, which is time-consuming. While Fourier slice photography \cite{Ng2005Fourier} can perform refocusing faster, it needs an initial 4-D Fourier transform, which is again slow; also, the quality of the Fourier refocused images are less than that of the slower shift-and-sum method because that discrete Fourier and inverse Fourier transform on the same data is not lossless. Also, both Fourier slice method and the shift-and-sum method are predicting each refocused image by interpolating some of the missing points; thus, they are simply estimates of the reality. Both methods sometimes have color and brightness prediction problems, which may need to be fixed by re-calibration. 

While \emph{Fourier slice photography} \cite{Ng2005Fourier} can be used to extract multiple refocused images from LFs faster than the shift-and-sum method by sacrificing a little quality, real-time refocusing of the LF, especially without time-consuming pre-processing, remains a pertinent challenge. Our goal is to address this problem by predicting the refocused images using a machine learning algorithm. By employing our model with a conventional desktop GPU, we can extract multiple focus-points from LFs in real-time with higher quality refocused images than both of the conventional refocusing methods. Real-time processing of the LFs can enable LF cameras to be used for detection tasks as well, possibly with better accuracy than obtained with conventional cameras. For example, LF cameras can be used for improved object detection because such algorithms can find objects from different focus points in LFs \cite{LF_object_detection2017}. In the case of conventional cameras, just a single focus point is available for detection, and blurred parts of the image need another capture with a different focus to be revealed. This problem can be solved by employing an LF instead of a conventional image.

In this paper, we introduce a residual network to estimate a set of 16 refocused images on different focal planes with refocusing parameters of $\alpha = \lbrace 0.125,0.250, 0.375, ...,2.0\rbrace$ \footnote{One can use any set of 16 refocusing parameters during the training phase.} from an LF with $7\times 7$ angular resolution. 
We show that our approach can predict refocused images in real-time, with superior color prediction, as compared to both of the conventional refocusing methods.

The code to accomplish our refocusing method will be publicly available upon acceptance. 
\section{Related Work}
\subsection{Refocusing techniques}
Here we review some of the previously developed refocusing methods in the literature.
Isaksnen \etal introduced some early algorithms to extract features like refocusing, synthetic apertures, and occlusion removals from LFs. Their refocusing algorithm is usually referred as brute force refocusing or reparametrization of LF \cite{Isaksen2000}. This method has time complexity of $O(n^4)$ and making it efficiently parallel is difficult. This method is slow, particularly because it needs to apply a homography to the whole LF for each requested focal plane. The shift-and-sum LF refocusing method was first introduced by Vanish \etal \cite{Vaish_SS} and later extended by Levoy \etal \cite{Levoy_SS}. This method first derives a mapping for the LF capturing device and the refocused images are extracted by warping the views in the LF using the calculated map. The quantized form of the shift-and-sum method was first derived by Ng \etal \cite{ng2005}. But the shift-and-sum method still has an asymptotic time-complexity of $O(n^4)$ and is not useful for real-time refocusing. \emph{Fourier slice refocusing} (FLR) was introduced by Ng \cite{Ng2005Fourier} in an attempt to speed-up the refocusing. Refocusing of the LF using FLR has the time complexity of only $O(n^2) \log n$, which is almost ideal for real-time refocusing. But the problem with this method is that it needs to calculate the four dimensional Fourier transform of each LF. In the discrete world, the 4-D fast Fourier transform's asymptotic time complexity is $O(n^4 \log n)$. Because of this pre-processing step, the LF cannot be refocused in real-time immediately after capture. Also, the refocused images extracted by the FLR method is inferior to the shift-and-sum method at some focus points because taking $FT^-1(FT(x))$ is not always equal to $x$ in discrete world. Nava \etal \cite{Nava2008} addressed the problem of FLR's quality by using generalization of the discrete Radon transform in the process. Their discrete focal stack transform produces better quality refocused images compared to FLR, but at the cost of reducing the number of possible refocusing focus points. In both shift-and-sum method and FLR, the camera array or the physically accurate conversion of the Plenoptic camera to camera array \cite{hedayati2020} should be on a single plane. If the cameras are placed on different planes the tilt-shift method can be used for refocusing \cite{tiltShift_1,Levoy_SS}. Other focus based features of LFs include multi-focus image extraction \cite{multi_focus} and super-resolved refocused images \cite{FourierSuperRes,Georgiev2011}. While these scenarios are not the focus of this paper, our method can be easily applied to these situations as well.

\subsection{ML Assisted Refocusing}
Depth estimation from single \cite{Srinivasan_2017_ICCV, hedayati_comp} or sparse LF views \cite{Kalantari2016} and LF reconstruction followed by using conventional refocusing techniques to demonstrate the quality of the LF have been thoroughly investigated. 
Deep learning has been used to extract refocused images using a single image as well \cite{Srinivasan_2018_CVPR,Sakurikar_2018_ECCV}. In stereo cameras, because of sparsity in the angular domain, the number of focal planes for refocusing are limited. Neural network have been employed to extract and enhance the refocused images for stereo cameras too 
 \cite{OccluDepthEst,MultiFusion2019}. However, we could not find any work in the literature targeting the particular problem of refocusing LFs captured by plenoptic cameras using machine learning.
\section{The Proposed Method}
In this section, our procedure for extracting refocused images from an LF is described. We have introduced a network for this purpose called RefNet. RefNet has the ability to extract 16 refocused images at different but static, predetermined focal planes.
\subsection{RefNet}
\begin{figure}
    \centering
    \includegraphics[width=0.35\textwidth]{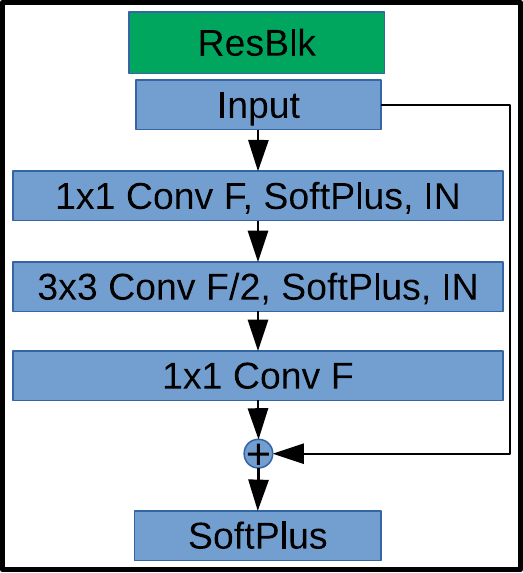}
    \caption{\label{fig:ResBlock} Structure of each ResBlock. Conv is short for 2-D convolution. The kernel size of each convolution is noted in the beginning of each cell. $F$ represents the number of filters of each convolution. IN is the instance normalization.}
\end{figure}

\begin{figure}
    \centering
    \includegraphics[width=0.47\textwidth]{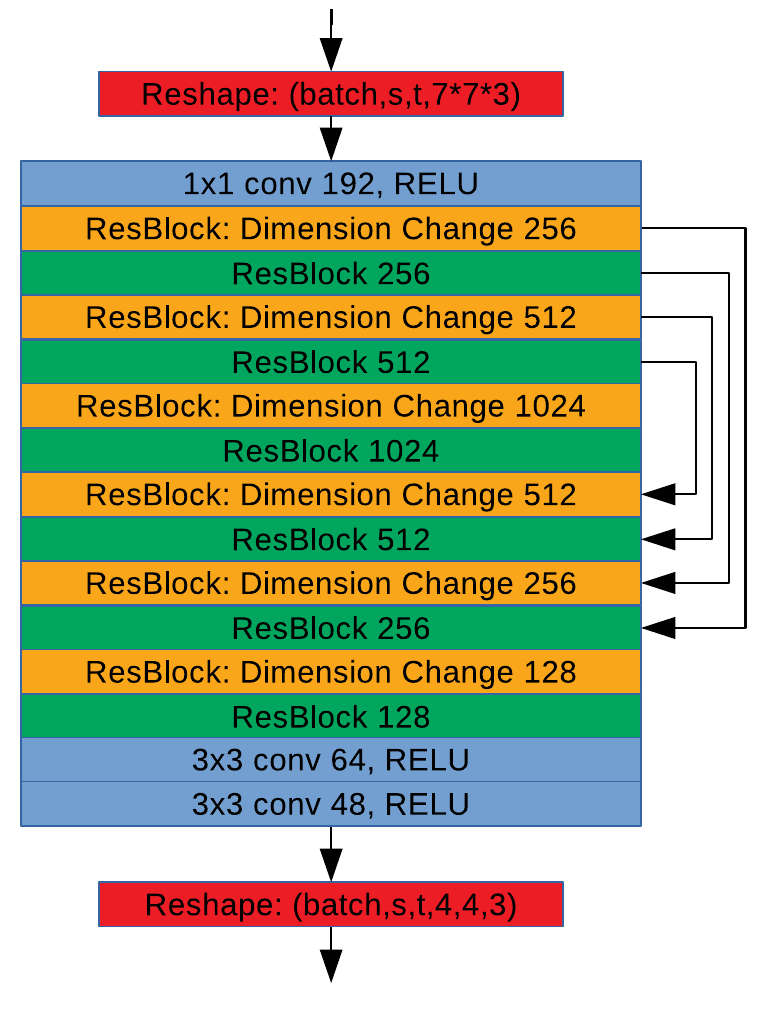}
    \caption{\label{fig:RefNet} The detailed structure of RefNet. First the LF is reshaped to $s,t,u\times v \times 3$, where in our experiment $u,v = 7$. We used a 2-D convolution with 192 filters as the first layer of the network followed by a 11 ResBlocks. Then with two additional convolution with RELU activation the final dimensionality  is reached. Finally, with a reshape, 16 estimated refocused image extracted.}
\end{figure}

Our network is constructed by employing 11 layers of modified ResNet50 \emph{bottelneck} building blocks \cite{Resnet}, or ResBlocks. These blocks, shown in Fig.~\ref{fig:ResBlock}, have been used for the main feature extraction. For filter dimensionality change, 2-D convolutions with $(1\times1)$ kernels have been used. The overall structure of the network is depicted in Fig.~\ref{fig:RefNet}.

Fig.~\ref{fig:ResBlock} illustrates that each ResBlock is composed of a 2-D convolution with kernel size of $(3\times3)$ and $F/2$ channels sandwiched between two 2-D convolution layers with $(1\times1)$ kernels and $F$ channels. Each of the first two convolutions are followed by a \emph{Softplus} activation function and an instance normalization. The input has a skip connection to the last convolution and then a Softplus activation is applied to the summed result. If the number of channels of the input differs from $F$, it will go through an additional 2-D convolution with $(1\times1)$ kernel and $F$ channels, then, it's result will be added to the last convolution.

For a detailed illustration of the RefNet structure, see Fig.~\ref{fig:RefNet}. First the LF is reshaped to $s,t,u\times v \times 3$. We used a 2-D convolution with $192$ channels as the first layer of the network with an RELU activation; this is followed by $11$ ResBlocks. The network then has two additional 2-D convolution layers with RELU activation. Finally, after a reshape, 16 refocused images are estimated as the network output.

The RefNet is trained in supervised manner. We use a weighted loss function for the purpose of training. The training output labels are the refocused images extracted from the LFs in our learning set by either the Fourier slice or shift-and-sum methods. Next we describe the loss function.

\subsection{Loss function} 
Our loss function is a weighted sum of the \emph{mean squared error} (MSE) and $\ell_1$-norm between the predicted and true images (as provided by the conventional reconstructions), and also the image quality losses as measured by \emph{structural similarity index metric} (SSIM) \cite{SSIM} and \emph{peak signal to noise ratio} (PSNR).

The first component of the loss function the appearance matching loss \cite{apearanceMatching}, which combines the SSIM with the $\ell_1$-norm,
\begin{equation}\label{eq:psi_1}
    \psi_1(I_p,I_g) = \beta\frac{1-SSIM(I_p,I_g)}{2} + (1-\beta)\left\|I_p-I_g\right\|_1,
\end{equation}
where $I_p$ and $I_g$ are the predicted and ground truth refocus images, respectively, and $\beta$ is a tuning parameter. The second component is the inverse PSNR,
\begin{equation}\label{eq:psi_2}
    \psi_2(I_p,I_g) = \frac{1}{PSNR(I_p,I_g)}.
\end{equation}
These two components are then added to MSE to form the final version of the loss function,
\begin{equation}\label{eq:loss}
    L = MSE(I_p,I_g) + \psi_1(I_p,I_g) + \gamma \psi_2(I_p,I_g).
\end{equation}
The tuning parameter $\gamma$ allows the magnitude of the inverse PSNR in $\psi_2$ to be tuned relative to the quantities in MSE and $\psi_1$, which we found to be helpful in overall performance. We provide the values of $\beta$ and $\gamma$ we used in our experiments in Section \ref{sec:exp}.

\subsection{Metrics}
For quantitative evaluation of the predictions, we used the SSIM and PSNR metrics. The average PSNR and SSIM over all of the focus points of one LF are calculated as
\begin{align}\label{eq:MPS}
    MSSIM & = \frac{1}{M} \sum_u{SSIM([I_p]_u,[I_g]_u)},\\
    MPSNR & = \frac{1}{M} \sum_u{PSNR([I_p]_u,[I_g]_u)},
\end{align}
where $[I_p]_u$ represents the $u$th refocused image extracted by our model and $[I_g]_u$ is the respective ground truth. $M$ is the total number of extracted refocused images.


\section{Experiments}\label{sec:exp}
In this section, we describe our method's implementation details for the experiments. Then we use public LF data sets \cite{Srinivasan_2017_ICCV,STDS,Kalantari2016} to evaluate our method and investigate the impact of different parts of our network on the performance of our model.

\subsection{Data sets}
We have conducted our experiments over three publicly available data sets: \emph{Flowers} \cite{Srinivasan_2017_ICCV}, \emph{30 Scenes} \cite{Kalantari2016}, and
\emph{Stanford Lytro Light Field Archive}  \cite{STDS}. We chose these data sets because all three of them are captured by Lytro Illum cameras, which is the most widely used LF camera. We cropped the LF from these data sets to the size of $7\times7\times375\times540$ to have a consistent size and vignetting.
We chose $648$ specific LFs from the Flowers data set, which were chosen to span a diverse set of viewing angles and scenes of the flowers. We did this because we noticed a lot of the LFs of the flowers were captured from the same scene with minimal camera movement. Another reason was to balance the number of flower scenes with other categories to have a balanced training set. Overall, our training data set contains $1,101$ LFs: $648$ flowers, $100$ LFs from the \emph{30 Scenes}, and the $353$ scenes from the \emph{Stanford archive}. We used the $30$ LFs of the \emph{30 Scenes test set} to evaluate our results; note that these 30 LFs are typically used as a benchmark set for LF reconstruction and are not contained in the training data.

\subsection{Implementation details}
We used the Fourier slice theorem and shift-and-sum method to pre-process
the training and test sets. Two new data sets of refocused images were created by extracting $16$ refocused images with $\alpha=\lbrace 0.125,0.250, 0.375, ...,2\rbrace$ from each LF. During the training phase, the LFs and the corresponding refocused images were randomly cropped to $[192,192,7,7]$. This cropping technique was used as an augmentation to the training data to prevent over-fitting the network during the training phase. We used a batch size of 4 crops of each LF during a single epoch. RefNet has about $12.5$ million trainable parameters. The networks were trained by minimizing the loss function using the ADAM optimizer \cite{kingma2017adam} for 90 epochs in our experiments. Each epoch took approximately $8$ minutes to complete; hence, the total amount of time spend for training was $13$ hours. 
In the training phase, $\beta=0.65$ was empirically found to yield the best results. And $\gamma=500$ performed well for training the network.

We implemented our model with Tensorflow 2.3 in python 3.7 on a workstation with an Intel Xeon W-2223 3.60 GHz, 64GB DDR4 memory, and an NVIDIA Quadro RTX 5000.

\subsection{Performance}
We compared our model's result with shift-and-sum and the Fourier slice method. Note that none of these methods can be claimed to be the actual ground truth refocused images. To obtain actual ground truth images, physical cameras would have to be used to capture images at different focus points. But such data are not available; hence, we compared our results with these benchmark approaches and evaluated the results visually as well. First we compared the our results quantitatively using PSNR and SSIM metrics, then we provided the refocused images for qualitative comparison as well. Please look at the electronic version of the paper for better quality assessment of images. In our supplementary materials, the full results of refocusing on the whole test set is available. 

\begin{figure}
    \centering
    \includegraphics[width=0.45\textwidth]{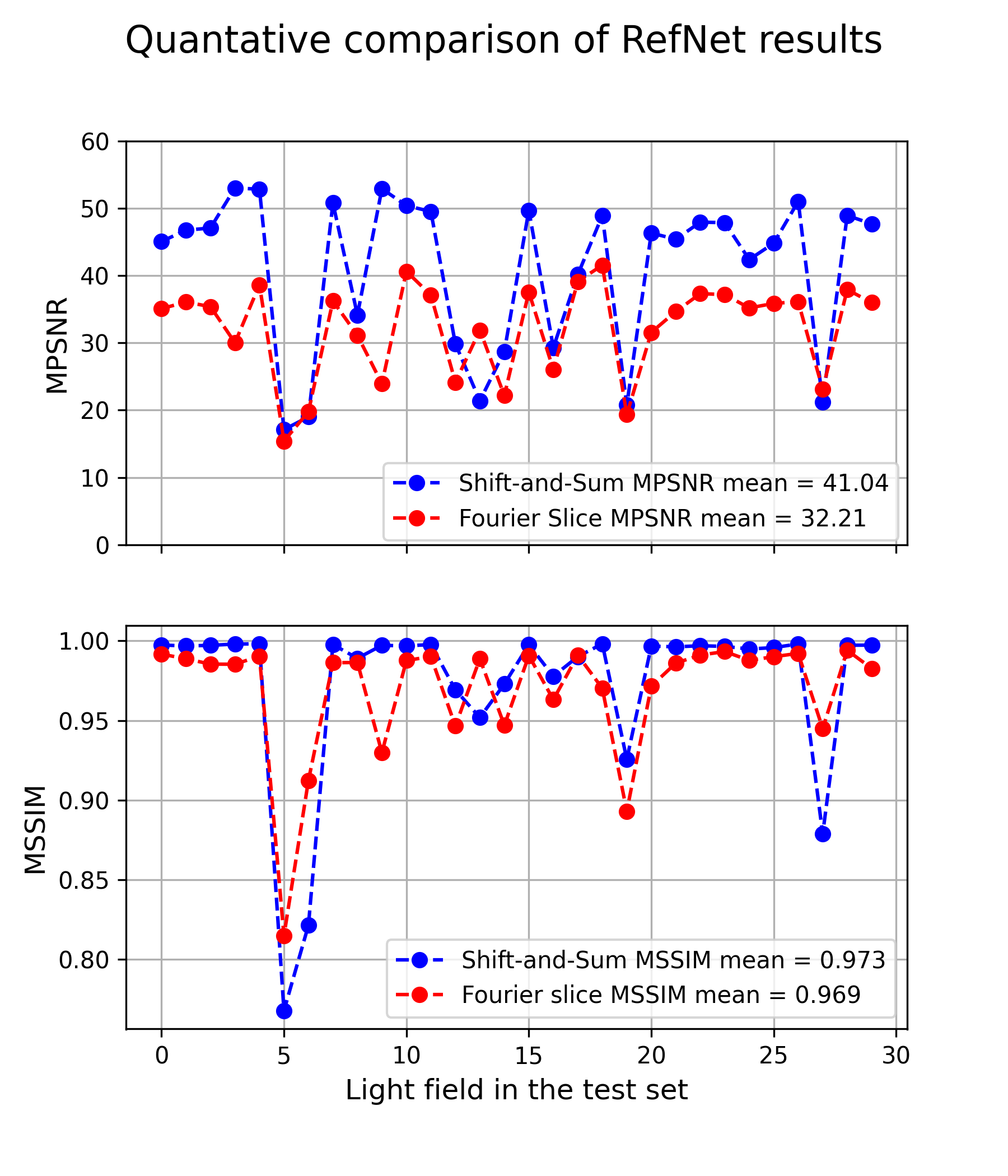}
    \caption{\label{fig:FSTvsSS} The MPSNR and MSSIM of the refocused images extracted from RefNet, compared to the shift-and-sum and Fourier slice methods. Blue points represent the MPSNRs and MSSIMs when the shift-and-sum refocused images are used as ground truth, while the red points represent the MPSNR and MSSIM of the RefNet refocused images predicted by the RefNet when it is trained on the Fourier slice method's refocused images. The set of refocused images for calculating the MPSNR and MSSIM are 16 refocused images with different focus points.}
\end{figure}

\begin{figure*}
    \centering
    \includegraphics[width=0.82\textwidth]{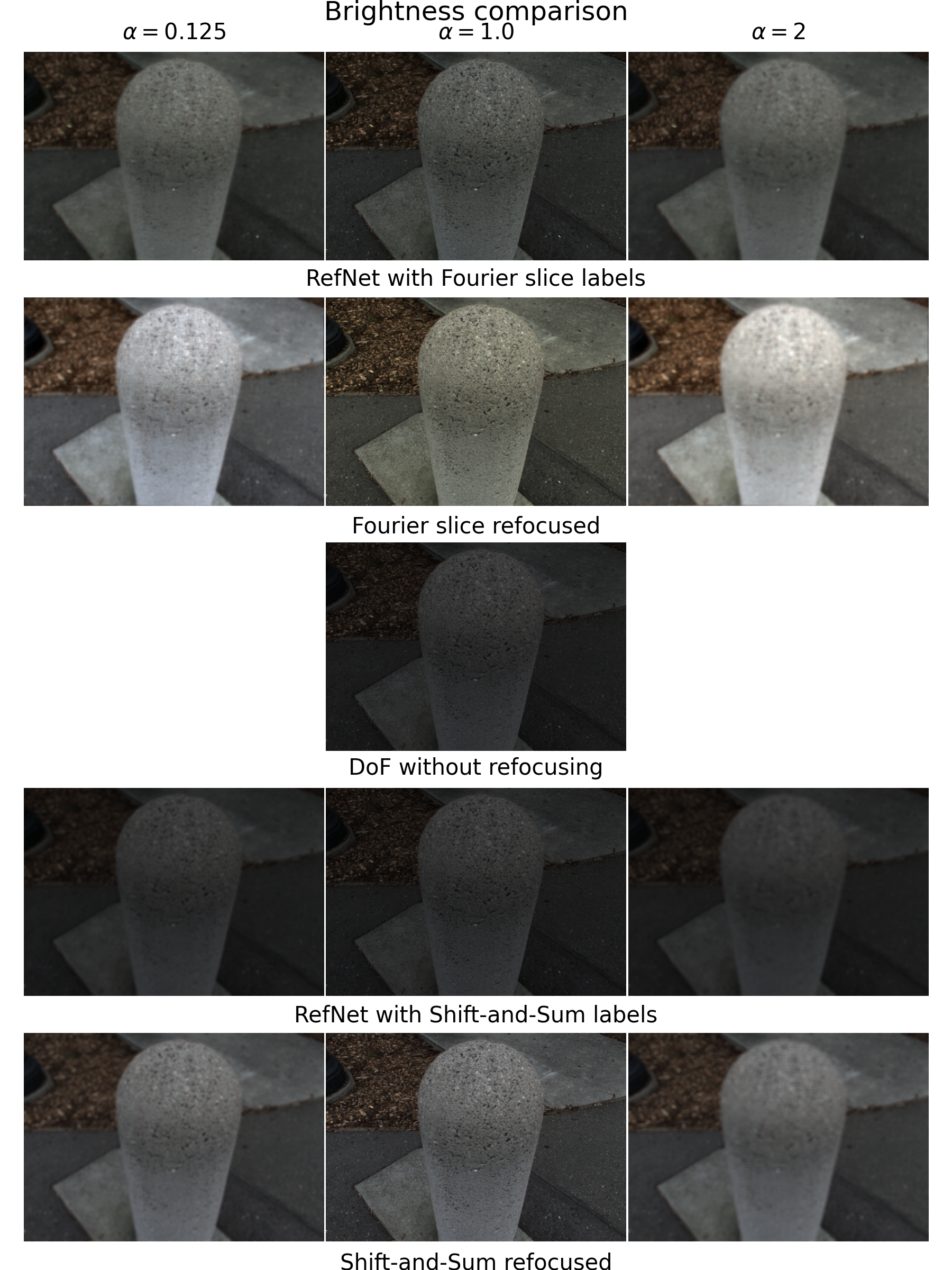}
    \caption{\label{fig:refComp_5} In this figure, three refocused images from each refocusing method are shown. Each row contains one method's refocused images with refocusing parameters of $\alpha=0.125$, $\alpha=1.0$ and $\alpha=2.0$ from left to right. In the middle, an unfocused DoF image is provided for visual comparison of the predicted brightness.
    From top to bottom, the used methods for refocusing are RefNet with Fourier slice ground truth, Fourier slice method, RefNet with shift-and-sum ground truth, and shift-and-sum method. The LF used for this figure is the fifth LF in the 30 Scenes data set \cite{Kalantari2016}.
    }
\end{figure*}

\begin{figure*}
    \centering
    \includegraphics[width=0.82\textwidth]{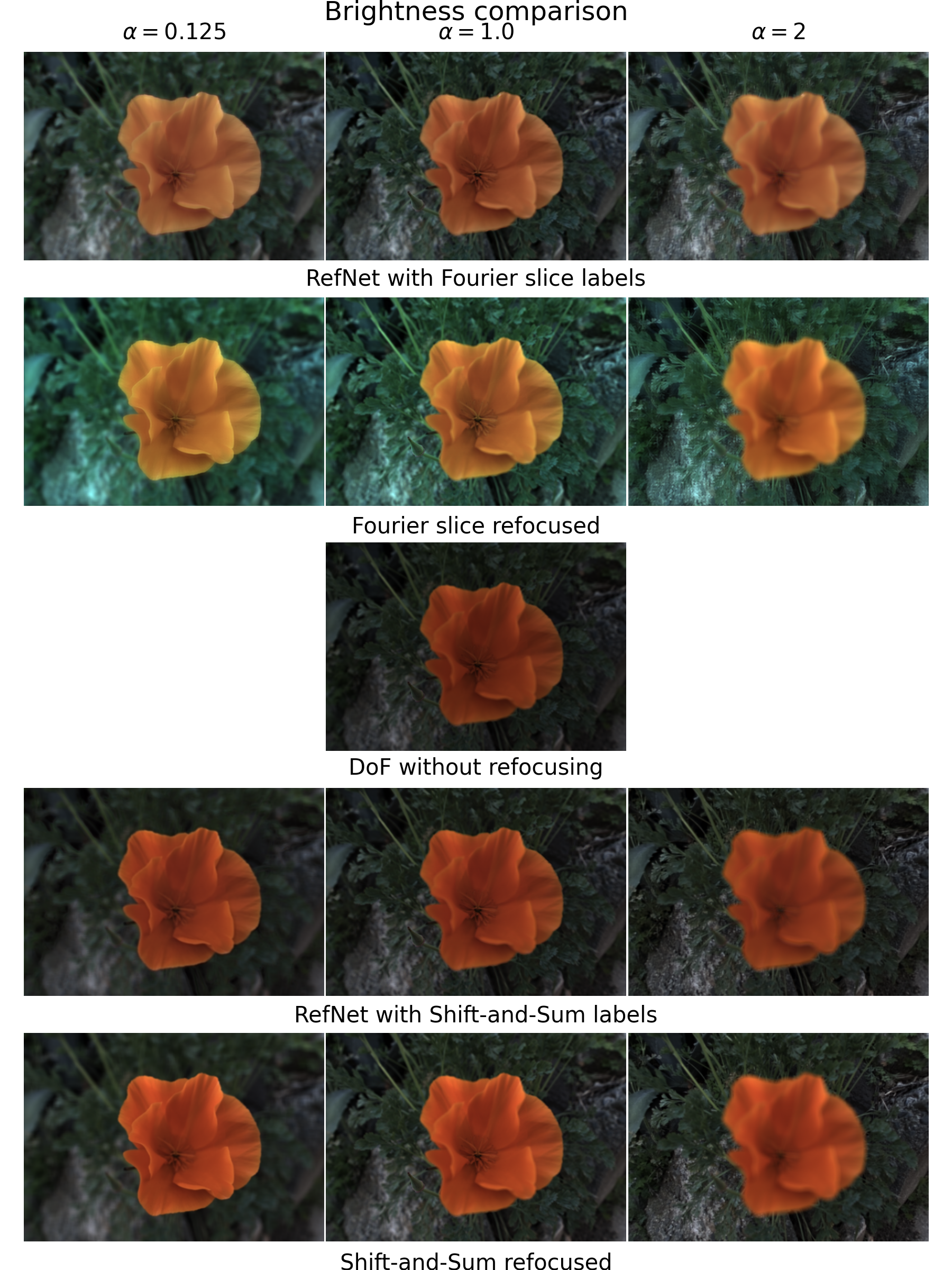}
    \caption{\label{fig:refComp_16} In this figure, three refocused images from each refocusing method are shown. Each row contains one method's refocused images with refocusing parameters of $\alpha=0.125$, $\alpha=1.0$ and $\alpha=2.0$ from left to right. In the middle, an unfocused DoF image is provided for visual comparison of the predicted colors.
    From top to bottom, the used methods for refocusing are RefNet with Fourier slice ground truth, Fourier slice method, RefNet with shift-and-sum ground truth, and shift-and-sum method. The LF used for this figure is the 16th LF in the 30 Scenes data set \cite{Kalantari2016}.}
\end{figure*}

The MSSIM and MPSNR values for the predicted refocused images of our model compared to the Fourier slice and shift-and-sum methods are plotted in Fig.~\ref{fig:FSTvsSS}. When the Fourier slice method is used as the ground truth for training, we can see that, most of the mean MSSIM values are more than $92.5\%$, and there are only two of them with a value less than $90\%$. We will show that RefNet's refocused images for these two LFs are better estimates compared to the FLR. For the case of MPSNR, the results are a bit worse than MSSIM in term of fluctuations. The majority of the MPSNR values are more than 30dB; however, at least seven trials show MPSNR values of less than 25dB. Similar behavior is observed for when the ground truth images are shift-and-sum refocused images. In this case, the average MSSIM and MPSNR is higher than the Fourier slice method, but there are still some trials that seem like outliers. By carefully looking at Fig.~\ref{fig:FSTvsSS}, we can see that some of the trials have high MSSIM and low MPSNR. This behavior---low MPSNR and high MSSIM---suggests a large color mismatch between the two sets but with a high structure similarity, which means they are focused roughly at the same focus point. Figs.~\ref{fig:refComp_5} and \ref{fig:refComp_16} demonstrate this impression. 

For qualitative comparison of the refocused images, we have compared refocused images extracted from RefNet trained with both the Fourier slice and shift-an-sum images for three different focus points, $\alpha=(0.125, 1.0, 2.0)$. See Figs.~\ref{fig:refComp_5} and \ref{fig:refComp_16}. The chosen LFs are number 5 and 16 in the test set. These LFs are selected because they are among the 5 trials in Fig.~\ref{fig:FSTvsSS} where MPSNR is around or below 20dB, indicating that the RefNet and ground truth are very different in either brightness or color. LF number 5 shows significant brightness differences and LF number 16 shows color differences. Similar figures that compare the refocused images for all of the 30 LFs in the test set are available in the supplementary materials.

In both of these figures, we can see that RefNet and the ground truth methods seemingly have the same DoF and focus point but they have distinction in color and brightness. Fig.~\ref{fig:refComp_5} shows that the refocused images from the Fourier slice method are significantly brighter than the DoF image without refocusing, while the brightness is not consistent among the three different focus points. For the shift-and-sum method, the brightness is consistent between different focus points but all of them are still brighter than the DoF image without refocusing while it has better predictions compared to the Fourier slice method. The RefNet trained on Fourier slice labels is performing almost similar to the shift-and-sum refocused images in brightness prediction and consistency. We can see that the RefNet with shift-and-sum labels has the best performance visually and the brightness of the RefNet refocused images are almost identical to the DoF image without refocusing. So, in terms of robustness in predictions, both trained RefNet networks are performing better than the Fourier slice method in terms of visual quality, while being able to produce these refocused images in real-time.

In Fig.~\ref{fig:refComp_16} we see that the RefNet approach is superior in terms of color reconstruction. The only difference is that the color of the predicted refocused images from RefNet that is trained by shift-and-sum labels is not identical to the DoF without refocusing, but it is the closest compared to other methods.

\subsection{Speed comparison}\label{sec:speed}
A truly fair speed comparison of FLR and RefNet is out of scope of this paper, as FLR requires a 4-D \emph{fast Fourier transform} (FFT) while the available GPU based FFTs are at most 3-D. Thus, we cannot implement the FLR method on GPU without tweaking the available tools and further optimizing for 4-D FFTs; though, we recognize this could be accomplished. Nor did we find an available GPU based implementation of the FLR. Hence, it was not possible to compare the speeds of the algorithms on the exact same GPU.
    
For the FLR method, the average time spent for 4-D FFT using NumPy's fftn \cite{harris2020array} over 1,000 samples, when the LF was loaded into CPU RAM, was $0.5137$ seconds over just the R channel. So the whole RGB LF would need $1.5411$ seconds just for the pre-processing step of the FLR. For RefNet, when the network is loaded into the GPU's memory and the LF is in CPU RAM (i.e., not yet loaded to GPU), a single forward propagation of one LF with all 3 RGB channels which extracts 16 refocused images takes $0.0209$ seconds. This is at least $73\times$ speed up compared to the pre-processing step of the FLR. If the LFs are grouped in batches, the speed gain of RefNet will be further improved. Using RefNet, a batch of 4 LFs can be processed, with 16 refocused images for each, in $0.046$ seconds. This is over $134\times$ faster than FLR. The significance of this speedup enables the possibility of real-time processing of LF videos concurrently with capturing. It is worth noting that the whole process of reading each LF from the hard drive to the final extraction of refocused images takes $\sim 0.2$ seconds per LF, which is still about an order of magnitude faster than just the pre-processing step of FLR, where, if the LF is being read from the hard drive, takes $1.85$ seconds.
    
\section{Conclusions and Future Work}
In this paper we have proposed a machine learning technique, called RefNet, to extract refocused depth of field images from a light field captured using a plenoptic camera. We demonstrated that by using our method, a light field can be refocused in real-time without pre-processing on conventional desktop GPUs. Furthermore, our method was demonstrated to be more visually pleasing in terms of color prediction when compared to the Fourier slice and shift-and-sum methods. RefNet is capable of extracting 16 refocused images with refocusing parameters of $\alpha=0.125,0.250,0.375,...,2.00$, though other refocusing parameter values could be trained for given proper training data. Possible future work includes exploring methods for training a network that can refocus light fields for any refocusing parameter, perhaps even using training data composed of other refocusing parameter values. Another avenue that can be explored is to extract multi-focus images from the light field using machine learning, which could enable 3-D object detection or reconstruction techniques.

{\small
\bibliographystyle{ieee_fullname}
\bibliography{refs}
}

\end{document}